\magnification=1200
\rightline{hep-th/9610207}
\bigskip
\centerline{\bf Knizhnik-Zamolodchikov-Bernard equations as a quantization}
\centerline{\bf of nonstationary Hitchin system.}

\bigskip
\centerline{\bf D.~Ivanov}
\centerline{12-127 M.I.T. Cambridge MA 02139 USA;}
\centerline{L.~D.~Landau Institute for Theoretical Physics, Moscow 117940, Russia}
\centerline{e-mail: ivanov@landau.ac.ru}

\bigskip
\centerline{October 24, 1996.}

\bigskip
{\narrower
\sevenrm\baselineskip=7pt 
The KZB equations for conformal blocks of the WZNW theory
are written on the moduli space of holomorphic
principal bundles on the surface. They become the multi-time Schr\"odinger
equation for the nonstationary Hitchin system. 
{}From the known form of the equations we learn about the
covariance of quantization with respect to changes 
of the coordinate frame.
\smallskip}
\bigskip

The Knizhnik-Zamolodchikov-Bernard (KZB) equations were first derived
as the equations for conformal blocks of the Wess-Zumino-Novikov-Witten
(WZNW) theory on Riemann surfaces [KZ,B,L,I].
These equations also appear in many other mathematical problems like
quantization of the moduli space of flat connections [H2,AWDP] or
quantization of isomonodromic deformations [R]. Many of the
known intgerable systems are closely related to the KZB equations
(mostly at genus one) [N,O]. The general
nature of these equations requires a more unified picture of the underlying
structures. This paper is a small step in this direction. As a starting
point we take the KZB equations in the form previously derived in the
framework of the WZNW theory [I]. We observe that these equations can be
pushed down from the non-invariant ``twist language'' of D.~Bernard [B]
to the moduli space of holomorphic principal bundles. 
The KZB operators realize a quantization of the quadratic Hamiltonians
of the Hitchin system [H1], the equations being the Schr\"odinger equations
in the multi-time of complex moduli of the surface. Classically, this
system has a quadratic kinetic Hamiltonian with a ``space-time'' dependent
mass. Under the quantization, the space dependence of mass leads to the
``potential'' term in the KZB equations. The multi-dimensionality of time
appears to provide an additional rigidity to the quantization procedure,
limiting the possible choice of the connection involved in the KZB
equations. We also observe a covariance of the KZB equations with respect
to changing the coordinate frame in accordance with the classical equations
of motion. This covariance generalizes the usual Galilean invariance
of the free particle in the flat space, but has also special features
due to the space-time dependence of mass and to the multidimensionality 
of time.

\def\cC {{\cal C}}
\def\cM {{\cal M}}
\def\bC {{\bf C}}
\def\cG {{\cal G}}
\def\cL {{\cal L}}
\def\cE {{\cal E}}
\def\BA {{\bar A}}
\def\Sw {{S_{_{\rm WZNW}}}}
\def\DL {{\cal L}}
\def\DR {{\cal R}}
\def\cA {{\cal A}}
\def\Div {{\rm div\,}}

\bigskip
\centerline{1. Definitions and constructions.}
\bigskip

Let $\cC$ denote the moduli space of Riemann surfaces of a given genus $N$
without punctures. For a given Riemann surface $\Sigma \in \cC$ we
consider the moduli space $\cM_\Sigma$ of stable holomorphic principal bundles
over $\Sigma$ with a complex Lie group $G_\bC$
($G_\bC$-bundles). The dimension
of $\cC$ is $3(N-1)$ (when $N>1$), $\dim \cM_\Sigma=(N-1)\dim G$. We
think of $\cM_\Sigma$ as a fiber of a bundle $\cE_N$ over $\cC$. $\cE_N$ is
the moduli space of stable $G_\bC$-bundles over Riemann surfaces
of genus $N$ with the structure group $G_\bC$. We shall parametrize
$G_\bC$-bundles by $(0,1)$ forms $\BA(z,\bar z)$ on $\Sigma$
with values in the complex Lie algebra $\cG_\bC$. A holomorphic
section $g$ is defined as a solution to the equation
$$ \bar\partial_{_\BA}\, g\equiv(\bar\partial -\BA)g=0,
\eqno(1.1)$$
where $\bar\partial$ is the antiholomorphic derivative on $\Sigma$.
Different $(0,1)$ forms $\BA$ and $\BA^\prime$ define the same holomorphic
principal bundle if the corresponding solutions of (1.1) differ by a
global shift $h(z,\bar z)$: $g^\prime=hg$. In other words, the forms
$\BA$ and $\BA^\prime$ define the same bundle if and only if they differ
by a globally defined gauge transformation
$$ \BA=h^{-1}\BA^\prime h - h^{-1}\bar\partial h.
\eqno(1.2)$$

In this paper we shall work {\it locally} in $\cM_\Sigma$ and disregard
its global structure. 

The cotangent space to $\cM_\Sigma$ is naturally identified with the
space of $(1,0)$ $\cG_\bC$-valued zero modes of $\bar\partial_\BA$, i.e.
(1,0) forms $J$ obeying
$$\bar\partial J - [\BA,J]=0.
\eqno(1.3)$$
The pairing of $J$ to a tangent vector $\delta\BA$ is defined by
$$\langle J,\delta\BA\rangle=\int_\Sigma J\delta\BA .
\eqno(1.4)$$
Due to the condition (1.3) this definition is consistent with the
gauge freedom for $\BA$.

Relating this picture to the WZNW model we treat $\BA$ as the external
source and associate to a particular choice of $\BA$ the holomorphic
partition function on the Riemann surface $\Sigma$:
$$ \Psi [\BA]=\langle e^{\int_\Sigma j^a \BA^a} \rangle,
\eqno(1.5)$$
where $j^a$ is the WZNW holomorphic current [KZ].
{}From the Polyakov-Wiegmann identity for the WZNW action
$\Sw [g]$
$$ \Sw [gh] = \Sw [g] + \Sw [h] - 2\int_\Sigma
(\bar\partial h ) h^{-1} g^{-1} \partial g
\eqno(1.6)$$
it follows that for gauge equivalent forms $\BA$ and $\BA^\prime$
the partition functions are related by a geometrical factor only:
$$\Psi [\BA] = \Psi [\BA^\prime] e^{k S[\BA \to \BA^\prime]},
\eqno(1.7)$$
where $k$ is the level of the WZNW theory, $S[\BA \to \BA^\prime]$
is a ``transition'' WZNW action (expressed by certain integrals over
the surface and the bulk, independent of $k$ and containing no
functional integration). This shows that the WZNW conformal block
is a (local) section of a line bundle $\cL_k$ over $\cM_\Sigma$ which is
the $k$-th degree of the level-one bundle (Quillen's determinant bundle
[H2,Q]):
$$ \cL_k=\cL_1^k.
\eqno(1.8)$$

In Bernard's formulation of twisted WZNW theory the principal bundle was
determined by constant transition functions along A-cycles (Bernard used
twist elements from the compact Lie group $G$, but we extend his construction
to the complexified group $G_\bC$). In our present language this corresponds to
$\BA$ concentrated on A-cycles only. Outside A-cycles $\BA=0$, in the
vicinity of the A-cycle $A_i$
$$ \BA=(\bar\partial\theta_i)\theta_i^{-1},
\eqno(1.9)$$
where $\theta_i(z,\bar z)$ is the ``step function'' equal to 1 on the
left of $A_i$ and to $g_i$ (Bernard's twist) on the right of $A_i$.
Actually, as $\theta$ sharpens, the partition function $\Psi [\BA ]$
diverges and needs a regularization. It can be shown that
 Bernard's definition of the (regularized)
twisted partition function $\Psi (g_i)$ results in a regularization
again of the form (1.7): $\Psi (g_i)=\Psi [\BA ] \, \exp (kS)$. Therefore
this regularization may be absorbed in (1.7) as the choice of local 
trivialization of $\cL_k$ and does not affect the further discussion.

Differentiation with respect to the constant twists in Bernard's picture
may be viewed as differentiation on the moduli space $\cM_\Sigma$. Let
$\omega^a_{ib}(z;w_0)$ form the basis of holomorphic twisted 1-forms:
$$\bar\partial_{_\BA} \omega^a_{ib} (z)=0 \quad (z\ne w_0), \qquad
\oint_{A_j} \omega^a_{ib} (z) \, dz =\delta_{ij}\delta^a_b
\eqno(1.10)$$
with a first-order pole at $z=w_0$. (In this paper we shall keep the
notation of [I].) Let $\DL^{ia}$ denote the right-invariant derivative
along the $i$-th twist. Then the differential $d_{\cM_\Sigma}$ on the
moduli space $\cM_\Sigma$ looks like
$$d_{\cM_\Sigma}(z)=\omega^a_{ib}(z;w_0)\DL^{ib}
\eqno(1.11)$$
(which is a differential operator mapping functions on $\cM_\Sigma$
to (1,0) forms obeying (1.3), i.e. to 1-forms on $\cM_\Sigma$; it
is regular at $z=w_0$ and, moreover, is independent of $w_0$).

Finally, we add to this picture the partition function of the twisted b-c system,
described in [B,L,I]. There it has been argued that the equations take the
most natural form when written not for the ``bare'' partition function
$\Psi (g)$, but for the product $F(g)=\Psi (g) \Pi (g)$,
where $\Pi(g)=\sqrt{Z_{b-c}}$ is the square root of the twisted
b-c partition function. The b-c theory is constructed with fermionic
fields $b^a(z)$ and $c^a(z)$ with spin 1 and 0 respectively and obeying the OPE
$$ b^a(z) c^b(w) ={\delta^{ab}\over z-w} +{\rm reg. terms}.
\eqno(1.12)$$
The b-c current has the form
$$ j_{b-c}^a(z)=f^{abc}b^b(z) c^c(z)
\eqno(1.13)$$
($f^{abc}$ are the structure constants of $\cG$) and obeys the
Kac-Moody algebra at level $2h^*$.
Similarly to the WZNW part, we define the b-c partition function as
$$ \eqalign{&(Z_{b-c})^{a_1\dots a_{N\dim G}}
(z_1,\dots z_{N\dim G}) [\BA]= \cr
&\langle e^{\int_\Sigma j^a_{b-c} \BA^a}
b^{a_1}(z_1)\dots b^{a_{N\dim G}}(z_{N\dim G})\prod_b^{\dim G}c^b (w_0)
\rangle.}
\eqno(1.14)$$
We need the insertions of $b$ and $c$ fields to eliminate zero modes. Integration
of the inserted $b$ fields along the A-cycles would recover the twisted
partition function used in [B,L,I]. For our purpose we leave the insertions
uncontracted. Since the partition function (1.14) is antisymmetric in $(z_i, a_i)$,
is a zero mode of $\bar\partial_{_\BA}$ on $\Sigma\setminus \{ w_0\}$ as a function
of each variable $z_i$, and has only a simple pole at $z_i=w_0$, it should be 
viewed as a holomorphic voulme form on the moduli space 
$\cM_{\Sigma\setminus \{ w_0\} }$
of $G_\bC$-bundles over $\Sigma\setminus \{ w_0\}$.
Sometimes (in genus one) it is more appropriate to work with
$\cM_{\Sigma\setminus \{ w_0\} }$ instead of $\cM_\Sigma$. However,
at genus $N>1$ we may freely convert $Z_{b-c}[\BA]$ into a holomorphic volume form
on $\cM_\Sigma$ by integrating $\dim G$ out of $N\dim G$ variables $z_i$
around $w_0$. This will remove the singularity at $z_i=w_0$ and leave
$\, \dim \cM_\Sigma = (N-1)\dim G \,$ variables $z_i$ as ``subscripts'' of
the holomorphic volume form $Z_{b-c}[\BA]$ on $\cM_\Sigma$. If we
denote the bundle of holomorphic volume forms (the canonical bundle)  
on $\cM_\Sigma$ as $K=\Omega^{(\dim \cM_\Sigma, 0)}(\cM_\Sigma)$, then
$Z_{b-c}[\BA]$ is a (local) section of the bundle
$$ \cL_{2h^*}\otimes K.
\eqno(1.15)$$
One may observe that as a line bundle $K$ is equivalent to $\cL_{-2h^*}$ [H2]
and therefore (1.15) is a trivial bundle. Still we shall keep
the two factors of (1.15) separately, since they are realized via different
mechanisms: the first factor is due to the anomalous gauge dependence of
$Z_{b-c}[\BA]$, while the second one comes from $Z_{b-c}[\BA]$ being explicitly
a holomorphic volume form. Now $\Pi=\sqrt{Z_{b-c}}$ is a holomorphic section
of $ \cL_{h^*}\otimes K^{1/2}$, and the ``dressed''partition function $F=\Psi\Pi$
is a section of 
$$ \cL_{\kappa}\otimes K^{1/2},
\eqno(1.16)$$
where $\kappa=k+h^*$.

To summarize our construction, the ``dressed'' partition function is a (local)
section of the bundle (1.16) over $\cM_\Sigma$ which is in turn a fiber of
a bundle over $\cC$. Therefore, the vector spaces 
$\Gamma(\cL_{\kappa}\otimes K^{1/2})$ of (local) holomorphic 
sections of (1.16) are fibers of a vector bundle over $\cC$. 
The KZB equations is a flat connection on this bundle.

\bigskip
\centerline {2. Nonstationary Hitchin system.}
\bigskip

This section is aimed to review the classical Hitchin system [H1] and
recall how the classical limit of the KZB operators coincides with the
quadratic Hitchin Hamiltonian [N].

The Hitchin system is a Hamiltonian system on the moduli space $\cM_\Sigma$
introduced in the previous section. The Hamiltonians $H_\mu$ are functions
on the cotangent bundle $T^*\cM_\Sigma$ equipped with the usual symplectic
structure $\omega=dp \wedge dq$. The Hamiltonians $H_\mu$ are parametrized
by the tangent vector $\mu$ to the moduli space of complex structures $\cC$;
alternatively we may speak of a Hamiltonian $H(z)$ being a quadratic
holomorphic differential in $z$ (the usual coupling to the Beltrami
differential $\mu(z,\bar z)$ is assumed: $H_\mu=\int_\Sigma H(z)
\mu(z,\bar z)$ ).
By nonstationary system we understand the collection of Hitchin systems
for all possible surfaces $\Sigma\in\cC$ which are the different moments
of the multidimensional time. At each ``time'' $\Sigma$ we have its own
set of Hamiltonians $H_\mu$ which govern the motion of the system as
``time'' changes. If we had an identification of different time slices
$\cM_\Sigma$, we would have been able to launch the system move in the
multi-time $\cC$ with the time dependent Hamiltonian $H_\mu$. The 
arrangement of such an identification of the fibers $\cM_\Sigma$ will be
an important issue in our future discussion.

The Hamiltonians $H_\mu$ can be defined as a symplectic quotient of a
free Hamiltonian in a certain affine space (see [AWDP] for an
analogous procedure over the moduli space of flat connections). We start
with the space $\cA_\Sigma$ of all $\cG_\bC$-valued (0,1) forms
$\BA^a$ and the free quadratic Hamiltonian $H_\cA (z)=J^2(z)/2$ on
$T^*\cA$ ($T^*\cA$ consists of (1,0) forms $J^a(z)$ ).\footnote{*}{In 
general, the Hitchin system includes also the higher-order
Hamiltonians. The KZB equations correspond to the quadratic part of the
Hitchin system. A generalization of the KZB equations to include the
higher-order Hamiltonians is an interesting mathematical problem.}
The Poisson bracket of the Hamiltonians at different points $z_1$ and $z_2$
is zero, and these Hamiltonians are invariant under the gauge 
transformations (1.2).

When we take the symplectic quotient with respect to the action of the 
gauge group, the space $\cA_\Sigma$ reduces to the moduli space $\cM_\Sigma$,
its cotangent space $T^*_\BA \cA$ --- to the space of zero modes $J(z)$
(i.e. $J(z)$ obeying (1.3)). The  Hitchin Hamiltonian being given by the
same expression
$$ H(z)={1\over 2} J^2(z)
\eqno(2.1)$$ is now a holomorphic quadratic differential.
If we parametrize the moduli space $\cM_\Sigma$ by Bernard's twists
$\{g_i\}$, the tangent space $T_\BA \cM_\Sigma$ has a corresponding
basis $\DL^{ia}$ (at this point  $\DL^{ia}$ are not operators yet,
they will become operators after quantization) modulo the relation
$\sum_i \DL^{ia}-\sum_i \DR^{ia} \equiv \sum_i (1-g_i)^a_b \DL^{ib}=0$. 
$J(z)$ is a linear function on the cotangent space $T^*_\BA \cM_\Sigma$:
$$ J^a(z)=\omega^a_{ib}(z;w_0)\DL^{ib}.
\eqno(2.2)$$
Then the Hitchin Hamiltonian (2.1) exactly reproduces the symbol of the KZB
operator from [I]. If we understand the classical limit of the KZB operator
as when acting on a rapidly oscillating function, then the ``potential''
term is negligible (of higher order in ``Planck constant'', see the
subsequent section for more detail), while the kinetic term reduces to
its symbol.

Thus, we explicitly verified that the quadratic Hitchin Hamiltonian is the
classical limit of the KZB operator.

\bigskip
\centerline{3. Multi-time Schr\"odinger equation.}
\bigskip

The KZB equation (on a surface without punctures) [I]
$$ \Big( \partial_m(z)+{1\over\kappa}\big[ {1\over2} \DL^{ia}
\omega_{ia}^b(z;w_0)\omega_{jc}^b(z;w_0)\DL^{jc} + U(z) \big]\Big) F=0
\eqno(3.1)$$
plays the role of the Schr\"odinger equation for the Hitchin Hamiltonian
(2.1).\footnote{${}^1$}{Unlike the approach of [O] (specific for the 
genus $N=1$ case), we treat the KZB equations
as a {\it nonstationary\/} Schr\"odinger equation.}
Here $\partial_m(z)$ is the differential on the moduli space of complex
structures $\cC$, $\kappa^{-1}=(k+h^*)^{-1}=\hbar$ is the ``Planck 
constant'', $U(z)$ is the ``potential'' term of the KZB operator
derived explicitly in [I]. 

Returning to the classical system, the Hitchin Hamiltonian (2.1) 
describes a free particle in a curved space, i.e. it is purely kinetic:
$$ H_\mu={1\over2} p_A C^{AB}_\mu p_B,
\eqno(3.2)$$
where $p_A$ are the momenta of the particle, $C^{AB}_\mu$ is the inverse 
mass matrix. Notice that $C^{AB}_\mu$ is space-time dependent; therefore
if we naively quantize this Hamiltonian by replacing $p_A \to \hbar
\partial / \partial x^A$, there is an ambiguity in possible ordering
the space derivatives and the space-dependent  $C^{AB}_\mu$. Of course,
we wish the Hamiltonian operator to be self-conjugate. This leads us to the
result
$$ \hat H_\mu={\hbar^2\over2} {\partial\over\partial x^A} C^{AB}_\mu
{\partial\over\partial x^B} + \hbar^2 U(x,t),
\eqno(3.3)$$
where $U(x,t)$ comprises the ambiguity of quantization. This $U(x,t)$
is the ``potential'' term in the KZB operator, but not the real potential
in the Hamiltonian, since it is of order of $\hbar^2$ and vanishes in the
classical limit. In principle, this term should depend on the coordinate
frame where the quantization is performed, and in the next section we
see that this is indeed the case.

\bigskip
\centerline{4. Covariance with respect to coordinate change.}
\bigskip

To write the KZB connection on the bundle of conformal blocks in the
form of the KZB equation (3.1) we need to fix a ``cordinate frame''. This
amounts to specifying the following three strucutres:

\item{(i)} First, since the ``dressed'' partition function $F$
is the square root of a holomorphic volume form ($K^{1/2}$ factor in 
(1.16)), one needs a reference holomorhic volume form $\omega$ on
$\cM_\Sigma$ to treat $F$ as a function.

\item{(ii)} Similarly, the $\cL_\kappa$ factor in (1.16) requires
a choice of local trivialization (or, equivalently, gauge fixing).

Once (i) and (ii) are chosen, it fixes a local trivialization
of the bundle (1.16) and allows us to regard $F$ as a function on
$\cM_\Sigma$.

\item{(iii)} Finally, to write a partial differential equation (3.1)
we need to specify the meaning of partial derivatives $\partial_m(z)$.
This is equivalent to picking a flat connection on the fiber bundle
$\cE$, so that $\partial_m(z)$ denote a differentiation along this
connection. At the same time, such a connection identifies different 
fibers $\cM_\Sigma$, which attributes a meaning to the nonstationary
Hitchin system as a Hamiltonian system with a time-dependent Hamiltonian.
In Bernard's formulation [B] this connection was based on fixing the
twisting group elements. Two points in different fibers $\cM_{\Sigma_1}$ 
and $\cM_{\Sigma_2}$ were identified if they had the same twists $\{g_i\}$.
Then $\partial_m(z)$ corresponded to changing the complex structure
keeping the twisting elements fixed.

As soon as all (i) --- (iii) items are chosen, the equations for
conformal blocks can be written as partial differential equations (3.1).
The freedom of choosing the coordinate frame (i) --- (iii) implies two
important questions. First, are there any natural or preferrable
choices and restrictions for choosing the coordinate frame?
Second, are the equations covariant with respect to coordinate frame
change?

Of course, we wish that the equations keep the form of a connection with
a spectral parameter $\hbar=1/\kappa$, i.e.
$$ [\partial_\mu + \hbar (A_{KZB})_\mu]F=0,
\eqno(4.1)$$
where $A_{KZB}$ is a second-order differential operator independent of
the level $k$ (or, equivalently, of $\hbar$).
Bernard's choice of coordinate frame is based on parametrizing
the holomorphic principal bundles by constant $\cG_\bC$ twists on
A-cycles. Since this parametrization is not unique and,
moreover, different twists (not related by a global conjugation) may
correspond to the same bundle in $\cM_\Sigma$ thus defining two
different coordinate systems in the vicinity of the same point in 
$\cM_\Sigma$, we expect that the equations (3.1) possess a certain
covariance with respect to a coordinate change.

Change of the bundle trivialization (i) --- (ii) transforms
$F$ as
$$ F \to F^\prime=FRe^{\kappa S},
\eqno(4.2)$$
where $R=(\omega^\prime / \omega)^{1/2}$ describes the
change of the reference volume form (i), $\exp(\kappa S)$
gives the change of the gauge (ii). We require that all the 
solutions to the KZB equation
$$ [\partial_\mu + {1\over\kappa}(\Delta_\mu + U_\mu)]F=0
\eqno(4.3)$$
are mapped under (4.2) to the solutions to the KZB equation
in the new coordinate frame:
$$ [\partial^\prime_\mu + {1\over\kappa}
(\Delta^\prime_\mu + U^\prime_\mu)]F^\prime=0.
\eqno(4.4)$$
Here $\Delta_\mu$ and $\Delta^\prime_\mu$ are second-order
differential operators (their symbols are equal and coincide
with the quadratic Hitchin Hamiltonian (2.1), but first-order
terms generally differ, since  $\Delta_\mu$ and 
$\Delta^\prime_\mu$ are self-conjugate with respect to different
volume forms $\omega$ and $\omega^\prime$, see more discussion
below). $U_\mu$ and $U_\mu^\prime$ are the ``potential''
terms of the KZB operators which also depend on the
coordinate frame. The connections $\partial_\mu$ and
$\partial^\prime_\mu$ differ by a Lie derivative along
vector fields $v_\mu$ on $\cM_\Sigma$:
$$\partial^\prime_\mu=\partial_\mu+\DL_{v_\mu}.
\eqno(4.5)$$

In addition, we require that in the initial coordinate frame
the following two conditions are satisfied. First, the reference form
$\omega$ is invariant with respect to the connection
$\partial_\mu$:
$$\partial_\mu \omega =0.
\eqno(4.6)$$
Second, the quadratic differential operator $\Delta_\mu$
is self-conjugate with respect to the reference form $\omega$.
This is a reflection of the coordinate-independent fact that
the KZB operator is self-conjugate on the line bundle $K^{1/2}$
(where self-conjugacy is naturally defined). Both
these conditions are satisfied in Bernard's coordinate frame.
Furthermore, the covariance requirement will enforce these
conditions in the new coordinate frame.

By a straightforward substitution (4.2) and comparing
terms at equal powers of $\kappa$, we arrive at the
following restrictions for the new coordinate frame:
$$
\partial^\prime_\mu \log R +\Delta_\mu S=0,
\eqno(4.7a)$$
$$
U_\mu-U^\prime_\mu={1\over R} \Delta^\prime_\mu R,
\eqno(4.7b)$$
$$
\partial^\prime S + {1\over2}[[\Delta^\prime_\mu, S], S]=0,
\eqno(4.7c)$$
$$
\DL_{v_\mu}F+[[\Delta^\prime_\mu, S], F]=0\quad
{\rm for\ any\ function\ }F.
\eqno(4.7d)$$

Using the self conjugacy of $\Delta_\mu$ with respect to 
$\omega$ and (4.7d), the equation (4.7a) may be rewritten as
$$ 2\partial^\prime_\mu \log R + \Div v_\mu =0,
\eqno(4.8)$$
where $\Div v_\mu ={1\over\omega}\DL_{v_\mu}\omega$ is the
divergence of $v_\mu$ with respect to the volume form $\omega$.
Using (4.6) this implies that
$$ \partial^\prime_\mu \omega^\prime =0,
\eqno(4.9)$$
i.e. the reference form must be conserved by the connection
$\partial_\mu$ in the new coordinate frame.

The equation (4.7b) indicates that the ``potential'' term indeed depends
on the coordinate frame, which was predicted in Section 3.

Finally, (4.7c) and (4.7d) state that $S$ satisfies the
Hamilton-Jacobi equation for the classical Hitchin
system and that $v_\mu$ is the velocity corresponding to
the classical action $S$. The Hamilton-Jacobi equation
(4.7c) appears in the highest order in $\kappa=\hbar^{-1}$,
which implies an analogy with the quasiclassical
limit of the ordinary quantum mechanics.

The described procedure of changing to a classically moving
frame presents an interesting generalization of a Galilean
invariance of the Schr\"odinger equation for a free particle
in a flat space. This generalization however has two
features special for our problem. First, the space is not
flat (the inverse mass matrix is space-time dependent),
therefore a ``potential'' term of the order of $\hbar^2$
appears because of the ambiguity of quantization. This
term depends on the coordinate frame via (4.7b). Second, the
time is multi-dimensional. It makes the ``velocities''
$v_\mu$ linearly dependent (since by (4.7d) they are linear
combinations of the first space derivatives of $S$). This
linear dependence defines a certain integrable contact
structure on $\cE_N$ which restricts the possible directions
of classical motion.

To summarize the results of this section, we observed that
to write the KZB equation we need the coordinate frame
consisting of a flat connection on $\cE_N$ and a trivialization
of the bundle (1.16). The KZB equations are covariant with 
respect to certain changes of the coordinate frame, when
the connection and the trivialization are changed
consistently. The changes of the connection must obey the
classical equations of motion for the Hitchin system (with
respect to the initial connection), while the change of
trivialization is given by the classical action along the
new trajectories. This condition separates a spectial class
of distinguished coordinate frames (Bernard's twist
parametrization being one of them).  In these coordinate
frames the KZB equation takes the usual form of a flat
connection with the spectral parameter $1/\kappa$, which
may be viewed as a covariance of quantization
of the Hitchin system in different coordinate frames.

\bigskip
\centerline{5. Natural connections on $\cE_N$.}
\bigskip

In the previous section we observed how starting from a given coordinate 
frame (understood as the three ingredients (i) --- (iii)) to construct 
a whole family of coordinate frames in accordance to the classical 
motion of the (nonstationary) Hitchin system. If the KZB equations 
have the usual form (4.1) in the initial frame, then they preserve 
this form under coordinate changes within this family. It appears 
that not all coordinate frames admit a quantization of the Hitchin 
system of the form (4.1). First of all, we must require that the 
connection $\partial_\mu$ on $\cE_N$ satisfies the two (classical) 
conditions:
$$ [\partial_\mu , \partial_\nu]=0;
\eqno(5.1)$$
$$ \partial_\mu H_\nu = \partial_\nu H_\mu.
\eqno(5.2)$$
The first condition is the flatness of the connection. The second one 
requires that the connection is compatible with the Hitchin 
Hamiltonians $H_\mu$ (as quadratic forms on the cotangent bundle,
these Hamiltonians can be differentiated along the connection 
$\partial_\mu$). Only if the conditions (5.1) --- (5.2) are satisfied 
may we hope that the classical Hamiltonians $H_\mu$ can be quantized 
to the operators $(A_{KZB})_\mu$ in (4.1) satisfying the 
integrability conditions
$$ \partial_\mu (A_{KZB})_\nu = \partial_\nu (A_{KZB})_\mu
\eqno(5.3)$$
and
$$[(A_{KZB})_\mu, (A_{KZB})_\nu]=0.
\eqno(5.4)$$
Naively counting the number of variables and equations in the 
conditions (5.1) --- (5.2) suggests that at sufficiently high genus 
these conditions are redundant. Fortunately, we know that Bernard's 
connection $\partial_\mu$ (keeping the twists unchanged) satisfies 
(5.1) --- (5.2) and, furthermore, admits a quantization (5.3) --- 
(5.4). We conjecture that at sufficiently high genus the 
conditions (5.1) --- (5.2) fix the possible choice of connections 
$\partial_\mu$ downto the family constructed from Bernard's 
connection by classical evolutions as described in the previous 
section. This selects a family of distinguished (``natural'') 
connections on the bundle $\cE_N$.

Finally, we make a remark how this notion of natural connections can 
be extended to the case of lower genera (where conditions (5.1) --- 
(5.2) are not restrictive enough), using the idea of compactifying 
the moduli space of Riemann surfaces so that the moduli of surfaces 
of lower genera form the boundary of the moduli of surfaces of higher 
genera. A detailed construction of this compactification may be found 
elsewhere [PN]. Here we employ only its basic idea to construct the natural 
connections over the moduli of the surfaces without marked points 
(punctures) from the natural connections over the moduli of the same 
surfaces with marked points. Inclusion of marked points increases the 
dimension of the moduli space, and with sufficiently many marked 
points the conditions (5.1) --- (5.2) will fix the family of natural 
connections uniquely.

So far in this paper we considered only surfaces without marked 
points. It has been noted in [I] that the KZB equations on the 
surface with marked points can be obtained from the equations without 
marked points by contracting handles of the surface to singular 
``nodes''. Below we show how to include marked 
points in the above constructions in a way which appears as 
the limit of contracting handles of the surface. Namely, if we have a 
surface $\Sigma$ with marked points $z_1,\dots,z_M$, then instead of 
$\cM_\Sigma$ we consider $\cM_{\Sigma,\{z_i\}}$ which is the moduli 
space of the (0,1) form $\BA$ together with group elements 
$g_1,\dots,g_M$ (located at the points $z_1,\dots,z_M$) with respect 
to the gauge transformations
$$ \BA \mapsto h^{-1}\BA h - h^{-1}\bar\partial h,
\qquad g_i \mapsto g_i h(z_i).
\eqno(5.5)$$
Defined in such a way, this moduli space naturally forms a fiber 
bundle over $\cM_\Sigma$:
$$ \cM_{\Sigma,\{z_i\}} \to \cM_\Sigma.
\eqno(5.6)$$ 
The cotangent vectors are the zero modes of $\BA$ (in the sense of 
(1.3)) with first-order poles at punctures $z_i$, the pairing being
$$ \langle J, \delta (\BA, g_i) \rangle
=\int_\Sigma J \delta \BA + \sum_i \oint_{z_i} J g_i^{-1} \delta g_i.
\eqno(5.7)$$
The holomorphic conformal block is a function of the gauge field
$\BA$ and the group elements $g_i$ transforming as (1.7) under gauge 
transformations. The usual conformal blocks may be recovered by 
projecting onto different finite-dimensional representations of 
$G_\bC^M$. For example, the usual Knizhik-Zamolodchikov equation will 
look in our formalism as
$$\left( {\partial\over\partial z_i} + {1\over\kappa} \sum_{j\ne i}
{\DL^{ia} \DL^{ja}\over z_i-z_j} \right) F(g_1,\dots,g_M) =0,
\eqno(5.8)$$
where $F$ is a function of group elements $g_i$ invariant under 
simultaneous left multiplication of all $g_i$ by the same group 
element. The whole discussion of the Hitchin system and the 
coordinate covariance can be extended to the case of surfaces with 
punctures in a straightforward way.

Suppose now that we have a unique family of natural connections 
$\partial_\mu$ on the fiber bundle $\cE_{N,M}$ of the moduli 
$\cM_{\Sigma,\{z_i\}}$ for surfaces with punctures (up to classical 
motion, see the previous section). Then select from this family only 
those connections which preserve the fibers in the bundles (5.6).
This subfamily of connections can be pushed down to connections on
$\cE_N$ which will again satisfy (5.1) --- (5.2). Of course, this 
construction recovers the family induced by Bernard's connection 
$\partial_\mu$ and distinguishes these connections as natural on
$\cE_N$ (where the conditions (5.1) --- (5.2) may be insufficient).

In the above treatment many points lack rigor; thus we put it as a 
conjecture that {\it there exists a unique family of connections on 
the fiber bundles $\cE_{N,M}$ which satisfy (5.1) --- (5.2) and which 
produce connections from the same family at lower genera under 
degenerating the surface; parametrization by Bernard's twists defines 
connections from this family.}

\bigskip
\centerline{6. Conclusion.}
\bigskip

The approach of this paper is to learn about the quantization of the 
Hitchin system from the KZB equations previously derived in the WZNW
theory. A more direct way would be to start with the Hitchin system
and to set up a procedure to consistently quantize it. Such a theory
has been developed in the 
works on geometric quantization of the moduli spaces of flat 
connections [AWDP,H2]. According to a theorem of Narasimhan and 
Seshadri [NS], the moduli space of flat connections on the surface
$\cM_N$ is isomorphic to the moduli space $\cM_\Sigma$ of 
stable $G_\bC$-bundles. $\cM_N$ has a natural 
symplectic structure, and fixing a complex structure $\Sigma\in\cC$
provides a complex structure on $\cM_N$ and, therefore, a K\"ahler
polarization. The result of [AWDP,H2] is that the quantization of
$\cM_N$ is independent of this polarization. This is achieved by 
providing a projectively flat connection on 
holomorphic sections of a certain line bundle over $\cM_N$. It was
shown by Hitchin [H2] that this geometric quantization provides the
quantum operators for the Hamiltonians (2.1). We strongly believe in
the uniqueness of the quantization of the Hitchin system, then the
connection of [AWDP,H2] must coincide with the KZB connection. This
would provide a correspondence between the two apparently different
quantization procedures: the geometric quantization and the path 
integral approach. A dictionary should be developed to translate 
between the structures outlined in this paper and those appearing in 
the geometric quantization of flat conections. It would also be 
interesting to understand the geometric meaning of the ``natural''
connections on $\cE_N$ conjectured in Section 5.

\bigskip
\centerline{Acknowledgements.}
\bigskip

The author is indebted to A.~Losev for inspiring this research. Most
of this work has been done at ITEP (Moscow) with its friendly and stimulating
spirit. The author wishes to thank N.~Nekrasov, V.~Fock, M.~A.~Olshanetskii,
and A.~Kirillov~Jr. for many useful discussions.

\bigskip
\centerline{References.}
\bigskip

\item{[H1]} N.~Hitchin, Duke Math.~J. {\bf 54} (1987) 91.
\item{[H2]} N.~Hitchin, Comm.~Math.~Phys. {\bf 131} (1990) 347.
\item{[AWDP]}  S.~Axelrod, S.~Della Pietra and E.~Witten, 
            J.~Diff.~Geom. {\bf 33} (1991) 787.
\item{[N]} N.~Nekrasov, {\it Holomorphic bundles and
            many-body systems}, preprint PUPT-1534, hep-th/9503157.
\item{[O]} M.~A.~Olshanetskii, {\it Generalized Hitchin
            systems and Knizhnik-Zamolodchikov-Ber\-nard equation on elliptic
            curves}, preprint SISSA-125-95-EP, hep-th/9510143.
\item{[I]} D.~Ivanov, Int.~J.~Mod.~Phys. {\bf A10} (1995)
            2507, hep-th/9410091.
\item{[B]} D.~Bernard, Nucl.~Phys. {\bf B303} (1988) 77;
   \item{} D.~Bernard, Nucl.~Phys. {\bf B309} (1988) 145.
\item{[L]}  A.~Losev, {\it Coset construction and
            Bernard equations}, preprint CERN-TH-6215/91.
\item{[KZ]} V.~G.~Knizhnik and A.~B.~Zamolodchikov, 
           Nucl.~Phys. {\bf B247} (1984) 83.
\item{[R]} N.~Reshetikhin, Lett.~Math.~Phys. {\bf 26} (1992) 167;
   \item{} J.~Harnad, {\it Quantum isomonodromic deformations and the
           Knizhnik-Zamolodchikov equations}, preprint CRM-2890 (1994),
           hep-th/9406078;
   \item{} D.~A.~Korotkin and J.~A.~H.~Samtleben, {\it On the quantization
           of isomonodromic deformations on the trous}, preprint
           DESY-95-202, hep-th/9511087.
\item{[NS]} M.~S.~Narasimhan and C.~S.~Seshadri, Math.~Ann. {\bf 155}
           (1964) 69.
\item{[PN]} P.~Nelson, Phys.~Rep. {\bf 149} (1987) 337.
\item{[Q]} D.~Quillen, Funct.~Anal.~Appl. {\bf 19} (1985) 31.

\end